\documentstyle[twocolumn,prl,aps]{revtex}
\begin{document}                                                              
\draft
\title{Possibility of spontaneous parity violation in hot $QCD$}
\author{Dmitri Kharzeev$^a$, Robert D. Pisarski$^b$, 
and Michel H.G. Tytgat$^{b,c}$}

\bigskip
 
\address{
a) RIKEN BNL Research Center, Brookhaven National Laboratory,
Upton, New York 11973-5000, USA\\
b) Department of Physics, Brookhaven National Laboratory,
Upton, New York 11973-5000, USA\\
c) Service de Physique Th\'eorique, CP 225, Universit\'e Libre de Bruxelles,
bvd du Triomphe, 1050 Bruxelles, Belgium}
\date{\today}
\maketitle
\begin{abstract} 
We argue that for $QCD$ in the limit of a large number of colors, the axial 
$U(1)$ symmetry of massless quarks is effectively restored at the deconfining 
phase transition.  If this transition is of second order, metastable states in
which parity is spontaneously broken can appear in the hadronic phase.  These 
metastable states have dramatic signatures, including enhanced production of 
$\eta$ and $\eta'$ 
mesons, which can decay through parity violating decay processes such as 
$\eta \rightarrow \pi^0 \pi^0$, 
and global parity odd asymmetries for charged pions.  
\end{abstract}
\begin{narrowtext}

It may be possible to observe the phase transition(s) 
from hadronic to quark and gluon degrees of freedom through
the collisions of heavy nuclei at ultrarelativistic energies.  
In the region of central rapidity, the relevant phase transitions
are those at nonzero temperature; these phase transitions
can be studied by numerical simulations of lattice gauge theory.
At present, simulations indicate that for three colors
coupled to light quarks, there is at most one phase
transition, controlled by the chiral 
dynamics of the light quarks \cite{lattice}.
The order of the phase transition in $QCD$, in which
two flavors are very light, and one flavor not too heavy
(up, down, and strange), is still unsettled.

The nature of the chiral phase transition
depends crucially upon the dynamics of the axial $U(1)$ symmetry
of the light 
quarks\cite{ren1,ren2}.  Notably, for two massless flavors, 
if the axial $U(1)$ symmetry is not 
restored about the chiral phase transition,
then the transition can be of second order; if it is 
restored, the transition may be driven first order
by fluctuations.

There are two approaches to understanding the dynamical breaking
of the axial $U(1)$ symmetry.
The first assumes that the dominant fluctuations are semiclassical
instantons\cite{instanton}-\cite{shuryak}.  
The second is based upon the large $N$ limit of
an $SU(N)$ gauge theory\cite{largeN}-\cite{meggiolaro},
and assumes that the dominant fluctuations are not
semiclassical, but quantum.

At zero temperature, both approaches give a
reasonably successful phenomenology for the $\eta'$ mass and
related processes.
In this Letter we show that these two approaches give radically
different predictions at nonzero temperature.  
In instanton models of the hadronic vacuum\cite{instanton},
the topological susceptibility is essentially constant below the
phase transition, and only drops off {\it above} the phase transition.
We argue that at large $N$, the topological
susceptibility essentially
vanishes {\it at} the phase transition.  If the deconfining
phase transition is of second order, then the axial $U(1)$ symmetry
is dynamically restored as the phase transition is approached from below.
Under this assumption, using a nonlinear sigma 
model\cite{largeNphen1}-\cite{meggiolaro} we show that 
metastable states with spontaneous parity violation arise
in the hadronic phase, and would produce striking experimental signatures.

The large $N$ limit of $SU(N)$ gauge theories is believed
to be a reasonable approximation even for $N=3$\cite{largeN}.
We assume that confinement holds for all $N$, 
with the masses of mesons and glueballs of
order one as $N \rightarrow \infty$; interactions between
mesons and/or glueballs are suppressed by powers of $1/N$.

Holding the number of quark flavors fixed as $N \rightarrow \infty$,
at large $N$ the $\sim N^2$ gluons dominate the $\sim N$ quarks.
Taking the degeneracy of hadronic bound states to be of order one,
the gluonic free energy changes from $\sim N^0$ in the hadronic
phase, to $\sim N^2$ in the deconfined phase.  Thus 
the gluonic part of the free energy 
can be used to define the temperature of the transition, 
at $T=T_d \sim N^0$\cite{thorn,rplargeN}.  
We further assume that any other transitions in the theory also
occur at $T_d$.  Given the huge change in the free
energy, any other possibility seems baroque, at best. 

In the pure glue theory, the topological susceptibility 
$
\lambda_{YM}(T) \equiv \partial^2 F(\theta, T)/\partial\theta^2
= \int d^4 x \, \langle Q(x) Q(0) \rangle
$,
where $F(\theta,T)$ is the free energy, and the 
$\theta$ parameter is conjugate to the integral of
the topological charge density, 
$
Q(x) = (g^2/32\pi^2) tr
(G_{\alpha \beta} \widetilde G^{\alpha \beta})
$.
At zero temperature, the free energy reduces to the energy,
$F(\theta,0)=E(\theta)$.

Because $Q(x)=\partial_\alpha K^\alpha$,
where $K^\alpha$ is the (gauge variant) topological 
current, $\lambda_{YM}(T)$
vanishes order by order in perturbation theory.  
At zero temperature, Witten suggested that 
quantum fluctuations generate
a nonzero value, $\lambda_{YM}(0) \sim N^0$\cite{witten1}.
At high temperature, the theory is asymptotically free
and so weakly coupled, with electric fluctuations
suppressed by Debye screening.
Thus at high temperature, $\lambda_{YM}(T)$
is unequivocally calculable by semiclassical means,
using instantons\cite{affleck,gpy}.
With $g^2$ the gauge coupling, the instanton action is $8\pi^2/g^2$;
as $g^2 N$ is held fixed when $N\rightarrow \infty$, 
$\lambda_{YM}(T) \sim exp(- a N)$, $a = 8 \pi^2/(g^2 N)$.  
This naive picture was verified by Affleck in a soluble asymptotically
free theory, the $CP^N$ model in $1+1$ dimensions\cite{affleck}.  

Thus for gauge theories, $\lambda_{YM}(T)$ changes from 
$\sim N^0$ at low temperatures, to $\sim exp(- a N)$ --- which
at large $N$ is essentially zero --- at high 
temperature.  Appealing to simplicity, we assume
that this change happens at the deconfining transition,
$\lambda_{YM}(T) \sim 0$ for $T > T_d$ \cite{affleck,davis}.

How $\lambda_{YM}(T)$ vanishes as $T \rightarrow T_d^-$ depends
upon the order of the deconfining phase transition.  
If the deconfining phase transition is of first order, then
since all interactions at large $N$ are suppressed by $1/N$,
presumably $\lambda_{YM}(T)= \lambda_{YM}(0)$ for $T < T_d$.

While there is some evidence that the
deconfining phase transition is of first order
for all $N \geq 4$, this conclusion 
may be premature\cite{ohta}.
Following the conjecture of \cite{rdptyt}, we henceforth assume 
that the deconfining phase transition is of second order at large $N$.
This requires that the phase
transition is driven by a Hagedorn spectrum\cite{rplargeN}.
Adding $N_f \geq 2$ flavors of massless quarks, 
we assume that chiral symmetry is spontaneously broken
at zero temperature \cite{largeN}, and that 
the deconfining phase transition
forces chiral symmetry restoration at $T=T_d$.
Since we shall argue that 
$\lambda(T) \rightarrow 0$ as $T\rightarrow T_d^-$,
the chiral transition is driven first order
by fluctuations\cite{ren1,ren2}.  
At large $N$, however, since the mesonic couplings start out small, 
$\sim 1/N$, fluctuations can only drive
the chiral transition first order within a 
narrow critical region, $\sim 1/N$.  
Thus the latent heat for the chiral transition is $\sim N^0$,
and relative to the $\sim N^2$ for the gluons in the deconfining phase,
it is a very weakly first order transition.
Consequently, we let quantities associated with the chiral transition
vary with temperature, but only as in mean field theory.  Notably,
the pion decay constant $f_\pi$, which is $\sim \sqrt{N}$, is
assumed to decrease as $f_\pi(T) \sim (T_d - T)^{1/2}$ when
$T \rightarrow T_d$.

While the pure glue theory depends upon $\theta$, the
addition of massless quarks must cancel any $\theta$ dependence.
Witten showed that this happens at large $N$ by the appearance of 
a light meson, the $\eta'$ \cite{witten1}.
We generalize this to nonzero temperature, 
to estimate how $\lambda_{YM}(T)$ and the $\eta '$ mass
vanish as $T \rightarrow T_d^-$.  

At large $N$, at zero temperature
any gauge invariant correlator is saturated by the exchange
of single glueballs and mesons \cite{largeN}.  Normally this changes
in a thermal bath, due to scattering off states in the thermal distribution.  
The hadronic phase of large $N$ $QCD$, however, is ``cold'':
chiral symmetry is restored not at a scale set by the pion
decay constant, $f_\pi \sim \sqrt{N}$ (as in, say, the sigma
model at large $N$), but at a much lower temperature, given
by the deconfining transition, $T_d \sim N^0$.  
Thus we can use the same type of
arguments as at zero temperature, simply allowing any
quantity which enters to be temperature dependent.

We start by following Veneziano\cite{veneziano}, and define 
$\lambda_{\eta'}$ as 
the form factor between the topological current and the $\eta'$ meson,
$
\langle 0 | K^\alpha | \eta'\rangle = 
i (\sqrt{N_f/N}) p^\alpha  \lambda_{\eta'}(T) 
$,
with $p^\alpha$ the momentum of the $\eta'$ meson.  
This form factor is precisely analogous to the coupling of
$\pi^0$ to two photons.  Following \cite{anom},
to one loop order in a constituent quark model, the (anomalous)
coupling of the $\eta'$ to two, or
indeed any finite number of gluons, vanishes as chiral
symmetry is restored, like
$\lambda_{\eta '}(T) \sim f_\pi(T)
\sim (T_d - T)^{1/2}$ as $T \rightarrow T_d^-$.  
Using Veneziano's relation, 
$m^2_{\eta '}(T) = N_f \lambda_{\eta '}^2(T)/N$, we find
the $\eta '$ mass vanishes as
$m^2_{\eta '}(T) \sim (T_d - T)$ when $T \rightarrow T_d^-$.  

We now use Witten's formula \cite{witten1} for the $\eta'$ mass,
$m^2_{\eta '}(T) = 4 N_f\lambda_{YM}(T)/f^2_\pi(T)$.
This relation shows that $\lambda_{YM}(T)$ and $f_\pi(T)$ must
vanish at the same point, but not much else.  Since Veneziano's
formula tells us how $m^2_{\eta '}(T)$ vanishes, though, we can
use this to deduce how the free
energy depends upon $\theta$ about $T_d$:
\begin{equation}
F(\theta,T) \sim (1 + c \, \theta^2) (T_d - T)^{2 - \alpha} \;\;
, \;\; T\rightarrow T_d^- \; ,
\label{e1}
\end{equation}
for some positive constant $c$, $|\theta| < \pi$.
That the $\theta$ dependence is only quadratic is characteristic of
large $N$ \cite{witten1}.
Then $\lambda_{YM}(T) = \partial^2 F(\theta, T)/\partial\theta^2
\sim (T_d - T)^{2 -\alpha}$ and
$m^2_{\eta '}(T) \sim (T_d - T)^{1 - \alpha}$ as $T \rightarrow T_d^-$.
Because the critical exponent
$\alpha \neq 0$, this does not quite agree with our estimate using
Veneziano's formula.  We trust (\ref{e1}), since the calculations
of \cite{anom} are only at one loop, and so
basically mean field.  Even so, as 
$\alpha \sim -.013$ \cite{ren1}, this difference is small, and does not alter
the conclusion that $\lambda_{YM}(T)$ and
$m_{\eta '}(T)$ vanish at the phase transition, with
$T_d$ and $\alpha$ independent of $\theta$.

Previously, Affleck\cite{affleck}
and also Davis and Mathieson\cite{davis}
argued that $\lambda_{YM}(T)$ vanishes when $T > T_d$;
our contribution is to estimate how it vanishes as $T \rightarrow T_d^-$
if the deconfining phase transition is of second order.
If the deconfining transition is of first order, then as
the hadronic phase is cold, presumably $\lambda_{YM}(T) = \lambda_{YM}(0)$ 
for $T < T_d$, at which point it drops discontinuously to zero.
(For an alternate view, with $T_d \neq T_\chi$,
see Meggiolaro\cite{meggiolaro}.)

An effective nonlinear sigma model which incorporates
the breaking of the axial $U(1)$ symmetry can be constructed 
\cite{largeNphen2}-\cite{meggiolaro}.  With
$U$ a $U(N_f)$ matrix satisfying $U^\dagger U = 1$,
the potential for $U$ is
\begin{equation}
V(U) = \frac{f_\pi^2}{2}
\left( tr\left( M(U + U^\dagger) \right) 
- a (tr \; ln \; U - \theta)^2 \right) \; ;
\label{e2}
\end{equation}
$M$ is the quark mass matrix.
When $M=0$, $m^2_{\eta '} \sim a$, so $a \sim
\lambda_{\eta'}^2/N$.  (Our $a$ = $a/N$ in 
\cite{largeNphen1}-\cite{meggiolaro}.)

Instanton processes are often modeled using a linear
sigma model with a field $\Phi$, by introducing a 
term $\sim e^{i \theta} det(\Phi)$ \cite{ren1}-\cite{instanton}; this 
term is well behaved in both the low and high temperature phases.
In contrast, in (\ref{e2}) the term 
$\sim (tr \, ln \, U)^2$, which breaks the axial $U(1)$ symmetry,
only makes sense in the low temperature phase,
and is singular if the vacuum expectation value (v.e.v.) of $U$ vanishes.
At large $N$, however, everything fits together: since
$a(T)\rightarrow 0$ as $T\rightarrow T_d^-$, 
there is simply no such term in the high temperature phase.
Moreover, at large $N$, even at zero temperature 
$\sim det(\Phi)$ must be dropped,
since it is inconsistent with the $\theta$ dependence \cite{witten2}.

Taking $M_{i j} = \mu_i^2 \delta^{i j}$,
any v.e.v of $U$ can be assumed to be diagonal, 
$U_{i j} = e^{i \phi_i} \delta^{i j}$; then
\begin{equation}
V(\phi_i) = f_\pi^2 \left(
- \sum_{i} \mu_i^2 \; cos(\phi_i)
+ \frac{a}{2} (\sum_i \phi_i - \theta )^2 \right) \; ,
\label{e3}
\end{equation}
which is minimized for
$
\mu_i^2 \; sin(\phi_i) + a (\sum \phi_i - \theta) = 0 \; .
$
Note that as  $\sum \phi_i$ arises from
$tr \, ln \, U$, it is defined modulo $2 \pi$.

All of the parameters in (\ref{e3}) are temperature dependent.
When $a=0$, the Goldstone boson masses 
$\mu_i^2 \sim m_i \langle \overline{q} q \rangle/f_\pi^2$,
with $m_i$ the current quark masses, and 
$\langle \overline{q} q \rangle$ the chiral order parameter.
When $M \neq 0$, there is no true critical point,
but we can use mean field theory to estimate that 
as $T \rightarrow T_d^-$, $f_\pi$ and $a$ decrease,
while the $\mu_i$ all uniformly increase:
$f_\pi^2(T) \sim a(T) \sim (T_d - T)$, 
$\langle \overline{q} q \rangle \sim (T_d - T)^{1/2}$, and
$\mu^2_i(T) \sim m_i/(T_d - T)^{1/2}$.
The solutions are independent of $f_\pi$,
and depend only upon the ratio $a/\mu_i^2$.  In mean field
theory this ratio is independent of flavor, and scales
as $a(T)/\mu_i^2(T) \sim (T_d - T)^{3/2}$.

Several authors have studied how the v.e.v's
for the $\phi_i$ change at zero temperature 
when $\theta \neq 0$ \cite{largeNphen1}-\cite{witten2}.  
Instead, we consider $\theta =0$, and follow Witten\cite{witten2}
to investigate metastable solutions at small $a$.
Related metastable states have been discussed by Shifman\cite{metastable}.

For a single flavor, the vacua are at 
$\phi = 0$, $\pm 2 \pi$, etc.
By balancing $\mu^2 sin(\phi)$ against $a \phi$, however, 
for small $a/\mu^2$ it is easy to show that there are other solutions
with $\phi \neq 0$.
These solutions have higher potential energy, and so
are local but not global minima.
Numerically, we find that the first metastable state
occurs when $a < a_{cr}$, with $a_{cr}/\mu^2 \sim .217$, and
$\phi_c \sim 4.493$; the $\phi$ field is massless about $\phi_c$.
As $a \rightarrow 0$, $\phi \rightarrow 2 \pi$,
which is equivalent to $\phi = 0$.
There is an infinite tower of metastable
states; we only consider that with lowest energy.

These metastable states are like regions with nonzero $\theta$,
and so spontaneously break $CP$ symmetry.
Under charge conjugation, $\phi \rightarrow + \phi$, 
while under parity, $\phi \rightarrow -\phi$.  Although there is a solution
at $- \phi$, 
when $- \phi$ does not differ from $+ \phi$ by a shift of $2 \pi$, 
parity is spontaneously violated.
This does not conflict with Vafa and Witten\cite{vafa},
who showed that at $\theta = 0$, parity is not spontaneously
violated in the $QCD$ vacuum.  Their theorem generalizes to the
thermodynamic minimum at nonzero temperature (although 
not at nonzero quark density), but it does not constrain metastable states.

The appearance of metastable states when $N_f \geq 2$ is somewhat
subtle.  To illustrate the basic point, we consider a two flavor model
in which $\mu_1^2 = 0$ and $\mu^2_2 = \mu^2$.  If the two
flavors decoupled\cite{witten2}, one might guess 
the solution $\phi_1 = 0$ and $\phi_2$ as for one flavor.
The equation of motion for $\phi_1$, however, forces 
$\phi_1 + \phi_2 = 0$: when $\mu_1^2=0$,
for any value of $a$ there is only
the trivial solution, $\phi_1 = \phi_2 = 0$, modulo $2 \pi$.

This example shows that there are no metastable states if any
quark mass vanishes.  This is natural: after all, there is also
no $\theta$ dependence when any quark mass vanishes, and these
metastable states are similar to regions with nonzero $\theta$.
Analogously, when all quarks have nonzero mass, 
metastable states only appear when
$a$ is small relative to the {\it lightest} quark mass.
Thus in $QCD$, for metastable states to occur
$a$ must be small relative not to the strange
quark mass, but to the up and down quark masses.
For this reason, $a(T)$ must become {\it very} small near the phase transition.

The potential of (\ref{e3}) can be used to obtain a qualitative
estimate.  Let $m_u$, $m_d$, and $m_s$ be the masses of
the up, down, and strange quarks.
The charged pions and the kaons are unaffected by the anomaly;
with $m^2_{\pi^\pm} \sim m_{u} + m_{d}$ and $m_K^2 \sim
m_{u,d} + m_{s}$, and assuming that $m_{u}= m_{d}/2$,
$m_{\pi^\pm} = 140 \, MeV$ and $m_K \sim 496 \, MeV$ give
$\mu_1^2 = (114 \, MeV)^2$, $\mu_2^2 = (161 \, MeV)^2$, 
and $\mu_3^2 = (687 \, MeV)^2$.  We take
Veneziano's\cite{veneziano} value of $a = (492 \, MeV)^2$, and
numerically diagonalize the mass matrix in (\ref{e3}) to
obtain $m_{\pi^0} \sim 139 \, MeV$, $m_\eta \sim 501 \, MeV$,
and $m_{\eta '} \sim 983 \, MeV$.  This is reasonably
close to the experimental values of $m_\eta \sim 548 \, MeV$,
and $m_{\eta '} \sim 958 \, MeV$.  

Taking the ratios of the $\mu_i^2$ as fixed, and varying
$a/\mu_1^2$, we studied numerically the appearance of
the lowest energy metastable state.
For the sake of discussion, we take the zero temperature $\mu_i^2$;
thus only the ratios of masses are believable, with all true
masses larger by some uniform factor, $\sim \mu_i^2(T)/\mu^2_i(0)
\sim (T_d - T)^{- 1/2}$.
The masses of the $\pi^0$, $\eta$, and $\eta'$ are read off
by diagonalizing the mass squared matrix obtained from (\ref{e3}).
The masses of the charged pions are
$m_{\pi^\pm}^2 = \mu_1^2 cos(\phi_1) + \mu_2^2 cos(\phi_2)$.
We ignore changes in the kaon masses; as $\phi_3$ is small, their
masses do not change much.

We find that there is a metastable solution when
$a/\mu_1^2 < .2467$, but it is unstable in the $\pi^0$
direction unless $a < a_{cr}$,
$
a_{cr}/\mu_1^2 \sim .2403 
$.
At $a_{cr}$, $\phi_1 \sim 4.47$, $\phi_2 \sim -.524
$, and $\phi_3 \sim -.028$;
the $\pi^0$ is massless at $a_{cr}$, while 
$m_{\pi^{\pm}} \sim 106 \, MeV$, 
$m_\eta \sim 150 \, MeV$, and $m_{\eta '} \sim 687 \, MeV$.
As $a \rightarrow 0$, the metastable state becomes equivalent
to the vacuum, as 
$\phi_1 \rightarrow 2 \pi$, $\phi_2$ and $\phi_3 \rightarrow 0$.
At $a=0$, $m_{\pi^0} \sim 114 \, MeV$, $m_{\pi^\pm} \sim 140 \, MeV$,
$m_{\eta} \sim 161 \, MeV$, and $m_{\eta'} \sim 687 \, MeV$.
In the thin wall approximation\cite{coleman}, 
the decay rate of the metastable 
state is $\Gamma \propto \exp(- F_c/T)$, where 
$F_c \sim (32 \sqrt 2 / 3) (\mu_1^3 f_\pi^2 / a^2)$.

Putting in the zero temperature values, in order for metastable
states to occur, near $T_d$ the
ratio of $a/\mu_1^2$ must be about $1\%$ of its value
at zero temperature.  
It is not clear if this is possible in
$QCD$, but of course this estimate is manifestly model dependent.
Since $F_c \sim 1/a^2$, at small $a$ the metastable states live
a very long time; thus in heavy ion collisions,
metastable states do not decay by bubble
nucleation; instead, as the hot phase cools, the
value of $a$ changes dynamically, and the metastable state rolls
smoothly into the true vacuum.

When $a$ becomes very small, there are several features common
to both the ground state and the metastable states.
First, the neutral Goldstone bosons are eigenstates not
of $SU(3)$, but of flavor\cite{ren1,largeNphen1}: at $a=0$, 
$\pi^0 \sim \overline{u} u$, 
$\eta \sim \overline{d} d$, and $\eta' \sim \overline{s} s$.
This generates maximal isospin violation: the neutral
pion is lighter than the charged pions, and so produced more
readily.  This effect is much stronger for the metastable states,
since the $\pi^0$'s are massless at $a_{cr}$, and so very light
for $a \sim a_{cr}$.
Similarly, the $\eta$ and $\eta'$ also become light
in both phases;
this is especially true for the $\eta$, as it
sheds all of its strangeness.  It is not clear how much
lighter the $\eta'$ becomes, given the overall increasing
mass scale of $\mu_i^2(T)$.
Light $\eta$ and $\eta'$ mesons are produced more readily \cite{kkm},
and can be observed
either directly, through $\gamma \gamma$ decays 
\cite{kkm}, or indirectly, through pion Bose-Einstein correlations 
\cite{vck}.

There are two types of experimental signatures special to the
formation of a parity violating phase.  The first is that
decays normally forbidden by parity are 
allowed (for a related phenomenon, see \cite{huang}).  
Kinematically, $\eta \rightarrow \pi^+ \pi^-$ is not allowed,
but $\eta \rightarrow \pi^0 \pi^0$ is.  
The processes $\eta' \rightarrow \pi^+ \pi^-$ and
$\eta' \rightarrow \pi^0 \pi^0$ are also allowed; however,
as the $\eta'$ is almost pure $\overline{s} s$, this is suppressed
by $\sim m_u/m_s$.

There are also global variables which are sensitive to the
dynamics of a parity violating phase.  
It can be shown that the interactions
of charged pions differ if there are regions with
$\phi_1$ and $\phi_2 \neq 0$, which change in either space or
time. This is similar to the propagation of charged
particles in a background magnetic field:
an $e^+e^-$ pair, produced back to back, are both deflected
in the same direction by a magnetic field.
A parity odd asymmetry could be observed by summing over all
$\pi^+\pi^-$ pairs in a given event,
\begin{equation}
{\cal P} = \sum_{\pi^+ \pi^-} \frac{[\vec{P}_{\pi^+} 
\times \vec{P}_{\pi^-}]\cdot \vec{z}}
{ |\vec{P}_{\pi^+}| |\vec{P}_{\pi^-}|};
\end{equation}
$\vec{z}$ is the beam axis of the collision, and
$\vec{P}_{\pi^\pm}$ are the pion momenta.
${\cal P}$ is like handedness in jet physics \cite{hand}.

These metastable domains might be of cosmological interest.
A region with $\phi_i \neq 0$ implies 
$\alpha_s G_{\alpha \beta} \widetilde G^{\alpha \beta}
\sim \lambda_{YM}(T) \sum \phi_i$ \cite{largeNphen1}; 
likewise, the coupling to electromagnetism should also
generate $\alpha \vec{E}\cdot\vec{B} \sim 
\alpha_s G_{\alpha \beta} \widetilde G^{\alpha \beta}$.
Thus if the entire universe fell into such a metastable domain,
it would generate a nonzero value for a cosmological magnetic field at the
time of the $QCD$ phase transition\cite{son}.

To summarize, in the limit of large $N$ the topological
susceptibility (essentially) vanishes in the deconfined phase.  If the
deconfining transition is of first order, then the susceptibility
is constant in the hadronic phase;
if of second order, the susceptibility vanishes as $T\rightarrow T_d$,
(1).  If the latter happens, metastable states in which parity is 
spontaneously broken can appear,
although in a nonlinear sigma model,
one must be very close to the phase transition for them to 
occur.  It is not clear if the large $N$ expansion is a good
guide to this physics when $N=3$; certainly at $N=3$, the
susceptibility will be nonzero in the high temperature phase.
Nevertheless, the large $N$ expansion provides a qualitative
guide against which other models can be tested.

We would like to thank M. Creutz, T. D. Lee, and D. T. Son for
comments and discussions.  This work was supported in part by DOE 
grant DE-AC02-98CH10886.

\end{narrowtext}
\end{document}